\documentclass[12pt]{article}

\usepackage[english]{babel}
\usepackage{amsmath,amsthm,amssymb,amsfonts,url}
\usepackage{mathptmx}      

\usepackage{color}

\newtheorem{example}{Example}
\allowdisplaybreaks
\newcommand{\Rey}{\hbox{Re}}

\begin{document}

\title{Approximately invariant solutions\\ of creeping flow equations}

\author{Matteo~Gorgone\\
\ \\
{\footnotesize Department of Mathematical and Computer Sciences,}\\
{\footnotesize Physical Sciences and Earth Sciences, University of Messina}\\
{\footnotesize Viale F. Stagno d'Alcontres 31, 98166 Messina, Italy}\\
{\footnotesize mgorgone@unime.it}
}

\date{Published in \textit{Int. J. Non-Linear. Mech.} \textbf{105}, 212--220 (2018).}

\maketitle

\begin{abstract}
In this paper, the  steady creeping flow equations of a second grade fluid in cartesian coordinates are 
considered; the equations involve a small parameter related to the dimensionless non--Newtonian 
coefficient. According to a recently introduced approach, the first order approximate Lie symmetries of the 
equations are computed, some classes of approximately invariant solutions are explicitly determined,
and a boundary value problem is analyzed.
The main aim of the paper is methodological, and the considered mechanical model is used to test the reliability of the procedure in a physically important application. 
\end{abstract}


\section{Introduction}
Lie theory of continuous transformations provides a unified and powerful approach for handling 
differential equations 
\cite{Ovsiannikov,Olver,Ibragimov:CRC,Baumann,BlumanAnco,Meleshko2005,Bordag,BlumanCheviakovAnco}. It is known that the knowledge 
of the Lie point symmetries admitted by ordinary differential equations allows for their order lowering and, possibly, reducing 
them to quadrature, whereas, in the case of partial differential equations,
symmetries can be used for the research of special (invariant) solutions of initial and boundary value problems. Also, the Lie 
point symmetries are important ingredients in the derivation of conserved quantities, or in the algorithmic construction of 
invertible point transformations linking different differential equations that turn out to be equivalent 
\cite{BlumanCheviakovAnco,Oliveri2010,Oliveri2012,GorgoneOliveri_RM,GorgoneOliveri_JGP2017}.

Unfortunately, any small perturbation of an equation usually destroys some important symmetries, and 
this limits the applicability of Lie group methods to concrete problems where equations involving terms 
of different orders of magnitude may occur. On the other hand, differential equations containing small terms 
are commonly and successfully investigated by means of perturbative techniques. 
To fill the gap, in the last decades, some \emph{approximate symmetry theories} have been proposed and widely 
applied to concrete models.

The first approach to approximate Lie symmetries is due to Baikov, Gazizov and Ibragimov \cite{BGI-1989}, who proposed to 
expand in a perturbation series the Lie generator in order to have an approximate generator.  They developed an elegant theory 
since all the useful properties of \emph{exact} Lie symmetries are adapted in the approximate sense.
Since its introduction, this approach has been applied to many physical models 
\cite{Wiltshire1996,BaikovKordyukova2003,DolapciPakdemirli2004,
IbragimovUnalJogreus2004,Wiltshire2006,GazizovIbragimovLukashchuk2010,
GazizovIbragimov2014}. 
Nevertheless, the expanded generator is not consistent with the principles of perturbative analysis 
\cite{Nayfeh} because the dependent variables are not expanded. This implies that in several examples 
the approximately invariant solutions obtained by this method are not the most general ones. 

Fushchich and Shtelen \cite{FS-1989} proposed a different approach. 
The dependent variables are expanded  in a series as 
done in usual perturbation theory; terms are then separated at each order of approximation, and a system of equations to be 
solved in a hierarchy is obtained. This resulting system 
is assumed to be coupled, and the approximate symmetries of the original equations are defined as the 
\emph{exact symmetries} of the equations obtained from separation. This approach has an obvious 
simple and coherent basis. \emph{Per contra}, a lot of algebra (especially for higher order perturbations) 
is needed; moreover, the basic assumption of a fully coupled system is too strong, since the equations at 
a level are not influenced by those at higher levels. In addition, there is no possibility to work in a 
hierarchy: for instance, if one computes first order approximate symmetries, and then searches for 
second order approximate symmetries, all the work must be done from the very beginning. 
Applications of this method to various  equations can be found, for instance, in the papers
\cite{Wiltshire2006,Euler1,Euler2,Euler3,Diatta}. Moreover, some variants \cite{DolapciPakdemirli2004,Valenti} of the Fushchich--Shtelen method have been proposed with the aim of reducing the amount of computations.

In particular, in \cite{DolapciPakdemirli2004} these two approaches have been compared, and a third method, which is a 
variant of Fushchich--Shtelen one, has been proposed, by removing the assumption of a fully coupled system. 
The involved  algebra is much less than that required by Fushchich--Shtelen method, and it is
possible to work in a hierarchy; nevertheless, the method is not general. An application of this approach, as well as a 
comparison of the results provided by the other approaches, has been given 
in \cite{DolapciPakdemirli2004} by considering the creeping flow equations of a second grade 
fluid. The conclusion of the authors is that their method is to be preferred either to the Baikov--Gazizov--Ibragimov approach or the Fushchich--Shtelen one. 

In a recent paper \cite{DSGO-2018}, a new approach to approximate symmetries has been proposed; the method is consistent with perturbative analysis, and inherits 
the relevant properties of exact Lie symmetries of differential equations. The main aim of this paper is to show that this new method provides reliable results, and avoids much of the weaknesses of the existing approaches. 
Therefore, the goal we want to pursue is concerned more with the methodological aspects of the used approach than with the solution of specific mechanical problems \cite{Bogus}.
In the method we apply, the dependent 
variables are expanded in power series of the small parameter as done in classical perturbative analysis; then, instead of 
considering the approximate symmetries as the exact symmetries of the approximate system (as done  in Fushchich--Shtelen method), 
the consequent expansion of the Lie generator is constructed, and the approximate invariance  with respect to the approximate Lie 
generator is introduced, as in Baikov--Gazizov--Ibragimov method. 
Of course, the method requires more computations than that required for determining exact Lie 
symmetries; nevertheless, a general Reduce \cite{Reduce} package (ReLie, \cite{Relie}), doing automatically all the needed work, 
is available.

This consistent theory allows to extend all the relevant features of Lie group analysis to an approximate context, \emph{i.e.}, 
it can be used to lower the order of ordinary differential equations as well as to compute approximately invariant solutions of 
partial differential equations \cite{DSGO-2018}.

In this paper, we apply this  consistent approach to the  creeping flow equations of a second 
grade fluid; the choice is motivated by the fact that this model has been analyzed in 
\cite{DolapciPakdemirli2004} 
by means of the different known approaches to approximate Lie symmetries whose results have been compared. 
Our aim is to show that the new consistent approach can effectively be applied in a concrete situation 
producing reliable results; the method proposed in  \cite{DSGO-2018} combines the ideas 
underlying the various approaches thus allowing to
take into account the principles of invariance under a continuous group of transformations and
perturbative analysis, so that it seems the more adequate one to deal with differential equations containing small terms. 

After determining the approximate Lie symmetries, by considering two different subalgebras of the
admitted Lie algebra of approximate symmetries, we determine explicitly the corresponding approximately 
invariant solutions. Also, we specialize one of these solutions by considering a boundary value problem for 
the model of a mud flow over a porous surface. It can be immediately seen that the approximate solution recovered for this boundary value problem turns out to be more general than the one determined in \cite{Pakdemirli2001}, where the  unperturbed model has been discussed.

The plan of the paper is as follows. In Section~\ref{sec:theory}, an overview of the new approach to
approximate Lie symmetries of differential equations is presented. In Section~\ref{sec:model}, the equations 
for the creeping flow of a second grade fluid are presented and briefly discussed: they involve small terms, 
the small parameter being the dimensionless non--Newtonian coefficient. Then the approximate Lie 
symmetries are computed. In Section~\ref{sec:solutions}, four classes of approximately 
invariant solutions are determined, and a boundary value problem discussed. Finally, Section~\ref{sec:conclusions} contains our conclusions.

\section{Approximate symmetry theories}
\label{sec:theory}

In this Section, a brief sketch of the approach to approximate Lie symmetries of differential equations developed in 
\cite{DSGO-2018} is given.

Let 
\begin{equation}
\Delta(\mathbf{x},\mathbf{u},\mathbf{u}^{(r)};\varepsilon)=0
\end{equation}
be a differential equation of order $r$, where $\mathbf{u}^{(r)}$ denotes the set of all derivatives of 
the dependent variables $\mathbf{u}\in U\subseteq\mathbb{R}^m$ with respect to the independent variables 
$\mathbf{x}\in 
X\subseteq\mathbb{R}^n$ up to the order $r$, involving a small parameter $\varepsilon$. 

If one looks for classical Lie point symmetries, in general it is not guaranteed that the infinitesimal 
generators depend on the parameter $\varepsilon$. Nevertheless, the occurrence of terms involving $
\varepsilon$ has dramatic effects 
since one loses some symmetries admitted by the unperturbed equation
\begin{equation}
\Delta(\mathbf{x},\mathbf{u},\mathbf{u}^{(r)};0)=0.
\end{equation}
In perturbation theory \cite{Nayfeh}, a differential equation involving small terms is often studied by 
looking for solutions in the form
\begin{equation}
\label{expansion_u}
\mathbf{u}(\mathbf{x},\varepsilon)=\sum_{k=0}^p\varepsilon^k \mathbf{u}_{(k)}(\mathbf{x})
+O(\varepsilon^{p+1}),
\end{equation}
whereupon the differential equation writes as
\begin{equation}
\Delta\approx \sum_{k=0}^p\varepsilon^k\widetilde{\Delta}_{(k)}\left(\mathbf{x},\mathbf{u}_{(0)},
\mathbf{u}^{(r)}_{(0)},
\ldots,\mathbf{u}_{(k)},\mathbf{u}^{(r)}_{(k)}\right)=0.
\end{equation}
Now, let us consider a Lie generator
\begin{equation}
\Xi=\sum_{i=1}^n\xi_i(\mathbf{x},\mathbf{u};\varepsilon)\frac{\partial}{\partial x_i}
+\sum_{\alpha=1}^m\eta_\alpha(\mathbf{x},\mathbf{u};\varepsilon)\frac{\partial}{\partial u_\alpha},
\end{equation}
where we assume that the infinitesimals  depend on the small parameter $\varepsilon$.

By using the expansion~\eqref{expansion_u} of the dependent variables 
only, we have the following expressions for the infinitesimals:
\begin{equation}
\xi_i\approx\sum_{k=0}^p\varepsilon^k \widetilde{\xi}_{(k)i}, \qquad \eta_\alpha\approx\sum_{k=0}
^p\varepsilon^k\widetilde{\eta}_{(k)\alpha},
\end{equation}
with
\begin{equation}
\begin{aligned}
&\widetilde{\xi}_{(0)i}=\xi_{(0)i}=\left.\xi_i(\mathbf{x},\mathbf{u},\varepsilon)\right|
_{\varepsilon=0},\qquad
&&\widetilde{\eta}_{(0)\alpha}=\eta_{(0)\alpha}=\left.\eta_\alpha(\mathbf{x},\mathbf{u},\varepsilon)
\right|
_{\varepsilon=0,}\\
&\widetilde{\xi}_{(k+1)i}=\frac{1}{k+1}\mathcal{R}[\widetilde{\xi}_{(k)i}],\qquad &&\widetilde{\eta}_{(k
+1)\alpha}=\frac{1}
{k+1}\mathcal{R}[\widetilde{\eta}_{(k)\alpha}],
\end{aligned}
\end{equation}
where the $\mathcal{R}$ is a recursion operator  defined as 
\begin{equation}
\label{R_operator_new}
\begin{aligned}
&\mathcal{R}\left[\frac{\partial^{|\tau|}{f}_{(k)}(\mathbf{x},\mathbf{u}_{(0)})}{\partial u_{(0)1}
^{\tau_1}\dots\partial 
u_{(0)m}^{\tau_m}}\right]=\frac{\partial^{|\tau|}{f}_{(k+1)}(\mathbf{x},\mathbf{u}_{(0)})}{\partial 
u_{(0)1}
^{\tau_1}\dots\partial u_{(0)m}^{\tau_m}}\\
&\phantom{\mathcal{R}\left[\frac{\partial^{|\tau|}{f}_{(k)}(\mathbf{x},\mathbf{u}_{(0)})}{\partial 
u_{(0)1}
^{\tau_1}\dots\partial u_{(0)m}^{\tau_m}}\right]}
+\sum_{i=1}^m\frac{\partial}{\partial u_{(0)i}}\left(\frac{\partial^{|\tau|} {f}_{(k)}(\mathbf{x},
\mathbf{u}_{(0)})}{\partial 
u_{(0)1}^{\tau_1}\dots\partial u_{(0)m}^{\tau_m}}\right)u_{(1)i},\\
&\mathcal{R}[u_{(k)j}]=(k+1)u_{(k+1)j},
\end{aligned}
\end{equation}
for $k\ge 0$,  $j=1,\ldots,m$, $|\tau|=\tau_1+\cdots+\tau_m$.
Thence, we have an approximate Lie generator
\begin{equation}
\Xi\approx \sum_{k=0}^p\varepsilon^k\widetilde{\Xi}_{(k)},
\end{equation}
where
\begin{equation}
\widetilde{\Xi}_{(k)}=\sum_{i=1}^n\widetilde{\xi}_{(k)i}(\mathbf{x},\mathbf{u}_{(0)},\ldots,\mathbf{u}
_{(k)})
\frac{\partial}{\partial x_i}
+\sum_{\alpha=1}^m\widetilde{\eta}_{(k)\alpha}(\mathbf{x},\mathbf{u}_{(0)},\ldots,\mathbf{u}_{(k)})
\frac{\partial}{\partial u_\alpha}.
\end{equation}

Since we have to deal with differential equations, we need to prolong the Lie generator
to account for the transformation of derivatives. This is done as in classical Lie group analysis of 
differential equations, \emph{i.e.}, the derivatives are transformed in such a way the contact 
conditions are preserved. 
Therefore, we have the prolongations
\begin{equation}
\begin{aligned}
&\Xi^{(0)}=\Xi,\\
&\Xi^{(r)}=\Xi^{(r-1)}+\sum_{\alpha=1}^m\sum_{i_1=1}^n\ldots\sum_{i_r=1}^n\eta_{\alpha,i_1\ldots i_r}
\frac{\partial}
{\partial \frac{\partial^r u_\alpha}{\partial x_{i_1}\ldots\partial x_{i_r}}},\qquad r>0,
\end{aligned}
\end{equation}
where
\begin{equation}
\eta_{\alpha,i_1\ldots i_r}=\frac{D \eta_{\alpha,i_1\ldots i_{r-1}}}{D x_{i_r}}-\sum_{k=1}^n\frac{D 
\xi_k}{D x_{i_r}}
\frac{\partial^r u_\alpha}{\partial x_{i_1}\ldots \partial x_{i_{r-1}}\partial x_k},
\end{equation}
along with the Lie derivative defined as
\begin{equation}
\frac{D}{Dx_i}=\frac{\partial}{\partial x_i}+\sum_{\alpha=1}^m\left(\frac{\partial u_{\alpha}}{\partial 
x_i}\frac{\partial}
{\partial u_{\alpha}}+\sum_{j=1}^n\frac{\partial^2 u_{\alpha}}{\partial x_i\partial x_j}\frac{\partial}
{\partial (\partial 
u_{\alpha}/\partial x_j)}+\cdots\right).
\end{equation}

Of course, in the expression of prolongations, we need to take into account the expansions of $\xi_i$, $
\eta_\alpha$, $u_\alpha$, and drop the $O(\varepsilon^{p+1})$ terms.

\begin{example}
Let $p=1$, and consider the approximate Lie generator
\begin{equation}
\begin{aligned}
\Xi &\approx \sum_{i=1}^n\left(\xi_{(0)i}+\varepsilon\left(
 \xi_{(1)i}+\sum_{\beta=1}^m\frac{\partial \xi_{(0)i}}{\partial u_{(0)\beta}}u_{(1)\beta}\right)\right)
 \frac{\partial}{\partial x_i}
\\
&+\sum_{\alpha=1}^m\left(\eta_{(0)\alpha}+\varepsilon\left(
 \eta_{(1)\alpha}+\sum_{\beta=1}^m\frac{\partial \eta_{(0)\alpha}}{\partial u_{(0)\beta}}u_{(1)\beta}
 \right)\right)
 \frac{\partial}{\partial u_\alpha},
 \end{aligned}
\end{equation}
where $\xi_{(0)i}$, $\xi_{(1)i}$, $\eta_{(0)\alpha}$ and $\eta_{(1)\alpha}$ depend on $(\mathbf{x},
\mathbf{u}_{(0)})$.
The first order prolongation is
\begin{equation}
\Xi^{(1)}\approx\Xi + \sum_{\alpha=1}^m\sum_{i=1}^n \eta_{\alpha,i}\frac{\partial}{\partial 
\frac{\partial u_\alpha}{\partial 
x_i}},
\end{equation}
where
\begin{equation}
\begin{aligned}
\eta_{\alpha,i} &= \frac{D}{D x_i}\left(\eta_{(0)\alpha}+\varepsilon\left(
 \eta_{(1)\alpha}+\sum_{\beta=1}^m\frac{\partial \eta_{(0)\alpha}}{\partial u_{(0)\beta}}u_{(1)\beta}
 \right)\right)\\
 &-\sum_{j=1}^n \frac{D}{D x_i}\left(\xi_{(0)j}+\varepsilon\left(
 \xi_{(1)j}+\sum_{\beta=1}^m\frac{\partial \xi_{(0)j}}{\partial u_{(0)\beta}}u_{(1)\beta}\right)\right)
 \left(\frac{\partial u_{(0)\alpha}}{\partial x_j}+\varepsilon \frac{\partial u_{(1)\alpha}}{\partial 
 x_j}\right),
\end{aligned}
\end{equation}
with the  Lie derivative now defined as
\begin{equation}
\frac{D}{Dx_i}=\frac{\partial}{\partial x_i}+\sum_{k=0}^p\sum_{\alpha=1}^m \left(\frac{\partial u_{(k)
\alpha}}{\partial x_i}
\frac{\partial}{\partial u_{(k)\alpha}}
+\sum_{j=1}^n\frac{\partial^2 u_{(k)\alpha}}{\partial x_i\partial x_j}\frac{\partial}{\partial (\partial 
u_{(k)\alpha}/\partial 
x_j)}+\cdots\right).
\end{equation}
Things go similarly for higher order prolongations.
\end{example}

The approximate (at the order $p$) invariance condition of a differential equation reads
\begin{equation}
\left.\Xi^{(r)}\Delta\right|_{\Delta\approx 0}\approx 0.
\end{equation}
In the resulting condition we have to insert the expansion of $\mathbf{u}$ in order to obtain the 
determining 
equations at the various orders in $\varepsilon$.

The Lie generator $\widetilde{\Xi}_{(0)}$ is always a symmetry of the unperturbed equations ($
\varepsilon=0$); the  \emph{correction}
terms $\displaystyle\sum_{k=1}^p\varepsilon^k\widetilde{\Xi}_{(k)}$ give the deformation of the symmetry 
due to the terms involving $\varepsilon$. 

It is worth of being remarked that not all the symmetries of the unperturbed equations are admitted as the zero--th terms of the 
approximate symmetries; the symmetries of the unperturbed equations that are the zero--th terms of the 
approximate symmetries are called \emph{stable symmetries} \cite{BGI-1989}. Moreover, 
if $\Xi$ is the generator of an approximate Lie point symmetry of a differential equation, then
$\varepsilon\Xi$ is a generator of an approximate Lie point symmetry too, but the converse is not true 
in general.

By the same arguments as in classical Lie theory of differential equations, it is easily ascertained that the approximate Lie 
point symmetries of a differential equation are the elements of an approximate Lie  algebra.

Approximate Lie symmetries of differential equations can be used to determine approximately invariant solutions by appending to 
the equations at hand the (approximate) invariant conditions. For example, for first order approximate Lie symmetries, the latter 
are
\begin{equation}
\begin{aligned}
&\sum_{i=1}^n\left(\xi_{(0)i}\frac{\partial u_{(0)\alpha}}{\partial x_i}+\varepsilon\left(\left(
 \xi_{(1)i}+\sum_{\beta=1}^m\frac{\partial \xi_{(0)i}}{\partial u_{(0)\beta}}u_{(1)\beta}\right)\frac{\partial u_{(0)\alpha}}
 {\partial x_i}+\xi_{(0)i}\frac{\partial u_{(1)\alpha}}{\partial x_i}\right)\right)\\
&\qquad -\left(\eta_{(0)\alpha}+\varepsilon\left(
 \eta_{(1)\alpha}+\sum_{\beta=1}^m\frac{\partial \eta_{(0)\alpha}}{\partial u_{(0)\beta}}u_{(1)\beta}
 \right)\right)\approx 0,
 \end{aligned}
\end{equation}
where $\alpha=1,\ldots,m$.

\section{The model and the admitted approximate Lie symmetries} 
\label{sec:model}
It is well known that the theory of Newtonian fluids can result inadequate in predicting the behavior of some fluids, so that constitutive relations for non--Newtonian fluid mechanics need to be considered. Several models have been proposed and examined to explain the nonlinear relationship between the stress and the velocity gradient; among them, a model which has gained much support both on theoretical and experimental reasons is that of second 
grade fluid. In this model, the constitutive equation for the stress tensor $T$ 
of an incompressible fluid reads
\begin{equation}
T=-p I+\mu A_1+\alpha_1 A_2+\alpha_2 A_1^2,
\end{equation}
where $p$ is the pressure, $\mu$ the viscosity, $\alpha_1$ and $\alpha_2$  material coefficients which may 
depend on the temperature, $A_1$ and $A_2$ are the first two Rivlin--Ericksen tensors 
\cite{Rivlin_Ericksen}.
The model of second grade fluids is compatible with termodynamics and allows  for stable solutions if 
\cite{Dunn_Fosdick}:
\begin{equation}\label{compat_cond}
\alpha_1+\alpha_2=0,\qquad \alpha_1\geq 0,\qquad \mu\geq 0.
\end{equation}
Under these hypotheses, the dimensionless field equations for incompressible second order fluids in 
vectorial form have been derived \cite{Dunn_Fosdick,Pakdemirli1992}.

To avoid the difficulties arising when the boundary has corners or concave regions,  a special orthogonal 
coordinate system \cite{Kaplun}, generated by the potential flow corresponding to an inviscid fluid, in 
which the streamlines and velocity potential lines are chosen as coordinate curves in the plane, can be 
used. In such a formulation, the equations of motion and the boundary conditions become independent of the 
shape of the body immersed into the flow.

Another hypothesis, useful in some applications, is that of slow motion;
in the case of  Newtonian fluids we have the Navier--Stokes equations that can be linearized, and exact 
solutions can be recovered under suitable boundary conditions. On the contrary,
in the case of a second grade fluid, written in a suitable curvilinear coordinate system, we are
led to the creeping flow equations, which are highly nonlinear. Either the Newtonian or the second grade 
creeping flow solutions, together with their respective theoretical features, have been discussed in 
several papers \cite{Tanner,Huilgol,Fosdick_Rajagopal,Rajagopal}.

In this Section, we consider the creeping flow equations of a second grade fluid, and compute the 
first order approximate Lie symmetries according to the new approach described in the previous Section.  
This model has been analyzed in \cite{Pakdemirli2001}, where the exact Lie symmetries have been  
determined, and in \cite{DolapciPakdemirli2004}, where the approximate Lie symmetries have been computed 
with the different approaches available in the literature. 

By considering the equations of motion for incompressible second order fluids
for a special curvilinear coordinate system $(\phi,\psi)$ \cite{Dunn_Fosdick,Pakdemirli1992} that we relabel as $(x,y)$, 
the steady plane creeping flow equations in cartesian coordinates read \cite{Pakdemirli2001}:
\begin{equation}\label{creep_eqs}
\begin{aligned}
\frac{\partial u}{\partial x}&+\frac{\partial v}{\partial y}=0,\\
\frac{\partial p}{\partial x}&-\frac{1}{\Rey}\left(\frac{\partial^2 u}{\partial x^2}+\frac{\partial^2 u}
{\partial y^2}\right)-\varepsilon\left(5\frac{\partial u}{\partial x}\frac{\partial^2 u}{\partial x^2}+
\frac{\partial u}{\partial x}\frac{\partial^2 u}{\partial y^2}+u\frac{\partial^3 u}{\partial x^3}+v
\frac{\partial^3 u}{\partial y^3}\right.\\
&+u\frac{\partial^3 u}{\partial x \partial y^2}+2\left.\frac{\partial v}{\partial x}\frac{\partial^2 v}
{\partial x^2}+\frac{\partial u}{\partial y}\frac{\partial^2 u}{\partial x \partial y}+\frac{\partial u}
{\partial y}\frac{\partial^2 v}{\partial x^2}+v\frac{\partial^3 u}{\partial x^2 \partial y}\right)=0,\\
\frac{\partial p}{\partial y}&-\frac{1}{\Rey}\left(\frac{\partial^2 v}{\partial x^2}-\frac{\partial^2 u}
{\partial x \partial y}\right)-\varepsilon\left(5\frac{\partial u}{\partial x}\frac{\partial^2 u}
{\partial x \partial y}-\frac{\partial u}{\partial x}\frac{\partial^2 v}{\partial x^2}-v\frac{\partial^3 
u}{\partial x \partial y^2}\right.\\
&+u\frac{\partial^3 v}{\partial x^3}-v\frac{\partial^3 u}{\partial x^3}+2\left.\frac{\partial u}{\partial 
y}\frac{\partial^2 u}
{\partial 
y^2}-\frac{\partial v}{\partial x}\frac{\partial^2 u}{\partial x^2}+\frac{\partial v}{\partial x}
\frac{\partial^2 u}{\partial y^2}-u\frac{\partial^3 u}{\partial x^2 \partial y}\right)=0,\\\end{aligned}
\end{equation}
where $u(x,y)$ and $v(x,y)$ are the velocity components in the $x$ and $y$ directions, respectively, 
$p(x,y)$ is the pressure, $\Rey$ is the Reynolds number, and $\varepsilon$ ($\varepsilon\ll 1/\Rey
$) is the dimensionless non--Newtonian coefficient, selected as the perturbation parameter. The resulting 
equations are non--dimensionalized and the coefficients entering the above equations are defined as 
follows:
\begin{equation}
\frac{1}{\Rey}=\frac{\mu}{\rho U L},\qquad \varepsilon=\frac{\alpha_1}{\rho L^2},
\end{equation}
where $L$ and $U$ are reference length and velocity, respectively, and the thermodynamic compatibility 
conditions \eqref{compat_cond} have been used (the interest reader 
may refer also to \cite{DunnRajagopal} for a detailed critical analysis of thermodynamical compatibility conditions of fluids of differential type). 
It is worth of being remarked that for creeping flows the inertial terms have been neglected, and for $
\varepsilon=0$ the classical form of the Navier--Stokes equations, in this coordinate system, can be 
recovered. 

By expanding the dependent variables, say
\begin{equation}\label{expand_variables}
\begin{aligned}
&u(x,y)=u_0(x,y)+\varepsilon u_1(x,y)+O(\varepsilon^2),\\
&v(x,y)=v_0(x,y)+\varepsilon v_1(x,y)+O(\varepsilon^2),\\
&p(x,y)=p_0(x,y)+\varepsilon p_1(x,y)+O(\varepsilon^2),
\end{aligned}
\end{equation} 
and using the consistent approach to approximate Lie symmetries \cite{DSGO-2018}, we obtain
that equations \eqref{creep_eqs} are approximately (at first order) invariant with respect to
the approximate Lie groups of point transformations generated by the following vector fields:
\begin{equation}\label{creep_sym}
\begin{aligned}
&\Xi_1=\frac{\partial}{\partial x}, \qquad \Xi_2=\frac{\partial}{\partial y},\qquad  \Xi_3=
\frac{\partial}{\partial p},
\qquad \Xi_4=\varepsilon\frac{\partial}{\partial x}, \qquad \Xi_5=\varepsilon\frac{\partial}{\partial 
y},\\
&\Xi_6=\varepsilon\left(u_0\frac{\partial}{\partial u}+v_0\frac{\partial}{\partial v}+p_0\frac{\partial}
{\partial p}\right),\qquad \Xi_7=\varepsilon\left(x\frac{\partial}{\partial x}+y\frac{\partial}{\partial 
y}-p_0\frac{\partial}{\partial p}\right),\\
&\Xi_8=x\frac{\partial}{\partial x}+y\frac{\partial}{\partial y}+(u_0+\varepsilon u_1)\frac{\partial}
{\partial u}+(v_0+\varepsilon v_1)\frac{\partial}{\partial v},\\
&\Xi_{9}=\varepsilon\left(\frac{\partial^2 f_1}{\partial y^2}\frac{\partial}{\partial u}-
\frac{\partial^2 f_1}{\partial x \partial y}\frac{\partial}{\partial v}+\left(f_2-\frac{1}{\Rey}
\left(\frac{\partial^3 f_1}{\partial x \partial y^2}+\frac{\partial^3 f_1}{\partial x^3}\right)
\right)\frac{\partial}{\partial p}\right),\\
\end{aligned}
\end{equation}
where $f_1=f_1(x,y)$ and $f_2=f_2(x)$ are arbitrary functions of the indicated arguments, along with the 
constraint
\begin{equation}\label{constraint}
\frac{df_2}{dx}-\frac{1}{\Rey}\left(\frac{\partial^4 f_1}{\partial x^4}+2\frac{\partial^4 f_1}{\partial 
x^2\partial y^2}+\frac{\partial^4 f_1}{\partial y^4}\right)=0.
\end{equation}
In order to determine approximately invariant solutions, in what follows we will solve the constraint 
\eqref{constraint} by 
choosing the \emph{ansatz}
\begin{equation}
f_1(x,y)= F(x) G(y)+H(x),
\end{equation}
whereupon three different cases may be considered:
\begin{enumerate}
\item[Case (i)]
\begin{equation}\label{b=0}
\begin{aligned}
&F(x)=a_3 x^3+a_4 x^2+a_5 x+a_6,\\
&G(y)=a_1y+a_2,\\
&H^{\prime\prime\prime}(x)=\Rey f_2(x)+a_7;
\end{aligned}
\end{equation}
\item[Case (ii)]
\begin{equation}\label{b>0}
\begin{aligned}
&F(x)=(a_3+a_5 x)\cos(bx)+(a_4+a_6 x)\sin(bx),\\
&G(y)=a_1\exp(by)+a_2\exp(-by),\\
&H^{\prime\prime\prime}(x)=\Rey f_2(x)+a_7;
\end{aligned}
\end{equation}
\item[Case (iii)]
\begin{equation}\label{b<0}
\begin{aligned}
&F(x)=(a_3+a_5 x)\exp(bx)+(a_4+a_6 x)\exp(-bx),\\
&G(y)=a_1\cos(by)+a_2\sin(by),\\
&H^{\prime\prime\prime}(x)=\Rey f_2(x)+a_7,
\end{aligned}
\end{equation}
\end{enumerate}
where $a_i$ $(i=1,\ldots,7)$ and $b$ are arbitrary constants.

\section{Approximately invariant solutions}
\label{sec:solutions}
Here, in order to construct approximately invariant solutions to \eqref{creep_eqs}, we consider two 
different one--dimensional subalgebras of the admitted  approximate Lie symmetries, \emph{i.e.},
\begin{equation}
\Xi_A=\kappa_1\Xi_3+\kappa_2 \Xi_6+\kappa_3\Xi_7+\Xi_8+\Xi_9,
\end{equation} 
and
\begin{equation}
\Xi_B=\Xi_1+\kappa_1\Xi_2+\kappa_2 \Xi_3+\kappa_3\Xi_4+\kappa_4\Xi_5+\Xi_9,
\end{equation}
where $\kappa_1$, $\kappa_2$, $\kappa_3$ and $\kappa_4$ are arbitrary constants. The first subalgebra essentially involves a stretching group of the independent and dependent variables, whereas the second one consists of the traslation of the independent variables and a non--uniform traslation of the dependent variables.

\subsection{Approximately invariant solutions with respect to $\Xi_A$}
Let us consider the following approximate generator 
\begin{equation}
\label{scale_oper}
\begin{aligned}
\Xi_A&=x(1+\varepsilon \kappa_3)\frac{\partial}{\partial x}+y(1+\varepsilon \kappa_3)\frac{\partial}
{\partial 
y}\\
&+\left(u_0+\varepsilon\left(u_1+\kappa_2 u_0+\frac{\partial^2 f_1}{\partial y^2}\right)\right)
\frac{\partial}{\partial u}\\
&+\left(v_0+\varepsilon\left(v_1+\kappa_2 v_0-
\frac{\partial^2 f_1}{\partial x \partial y}\right)\right)
\frac{\partial}{\partial v}\\
&+\left(\kappa_1+\varepsilon\left((\kappa_2-\kappa_3)p_0+f_2-\frac{1}{\Rey}
\left(\frac{\partial^3 f_1}{\partial x \partial y^2}+\frac{\partial^3 f_1}{\partial x^3}\right)
\right)\right)\frac{\partial}{\partial p}.
\end{aligned}
\end{equation}
The corresponding approximately invariant solutions are such that
\begin{equation}
\begin{aligned}
x(1+\varepsilon \kappa_3)\frac{\partial u}{\partial x}+y(1+\varepsilon \kappa_3)\frac{\partial u}
{\partial y}&=u_0+\varepsilon\left(u_1+\kappa_2 u_0+\frac{\partial^2 f_1}{\partial y^2}\right),\\
x(1+\varepsilon \kappa_3)\frac{\partial v}{\partial x}+y(1+\varepsilon \kappa_3)\frac{\partial v}
{\partial y}&=v_0+\varepsilon\left(v_1+\kappa_2 v_0-
\frac{\partial^2 f_1}{\partial x \partial y}\right),\\
x(1+\varepsilon \kappa_3)\frac{\partial p}{\partial x}+y(1+\varepsilon \kappa_3)\frac{\partial p}
{\partial y}&=\kappa_1+\varepsilon\left((\kappa_2-\kappa_3)p_0\phantom{\frac{}{}}\right.\\
&+\left.f_2-\frac{1}{\Rey}
\left(\frac{\partial^3 f_1}{\partial x \partial y^2}+\frac{\partial^3 f_1}{\partial x^3}\right)
\right),
\end{aligned}
\end{equation}
whereupon, insertion of \eqref{expand_variables}, and separation of the coefficients of different powers 
of $\varepsilon$, provide the system
\begin{equation}\label{eq_scale}
\left\{\begin{aligned}
x\frac{\partial u_0}{\partial x}&+y\frac{\partial u_0}{\partial y}=u_0,\\
x\frac{\partial u_1}{\partial x}&+y\frac{\partial u_1}{\partial y}+\kappa_3\left(x\frac{\partial u_0}
{\partial x}+y\frac{\partial u_0}{\partial y}\right)=u_1+\kappa_2 u_0+\frac{\partial^2 f_1}{\partial y^2},
\\
x\frac{\partial v_0}{\partial x}&+y\frac{\partial v_0}{\partial y}=v_0,\\
x\frac{\partial v_1}{\partial x}&+y\frac{\partial v_1}{\partial y}+\kappa_3\left(x\frac{\partial v_0}
{\partial x}+y\frac{\partial v_0}{\partial y}\right)=v_1+\kappa_2 v_0-
\frac{\partial^2 f_1}{\partial x \partial y},\\
x\frac{\partial p_0}{\partial x}&+y\frac{\partial p_0}{\partial y}=\kappa_1,\\
x\frac{\partial p_1}{\partial x}&+y\frac{\partial p_1}{\partial y}+\kappa_3\left(x\frac{\partial p_0}
{\partial x}+y\frac{\partial p_0}{\partial y}\right)=(\kappa_2-\kappa_3)p_0\\
&+f_2-\frac{1}{\Rey}
\left(\frac{\partial^3 f_1}{\partial x \partial y^2}+\frac{\partial^3 f_1}{\partial x^3}
\right).
\end{aligned}
\right.
\end{equation}
By considering the case (i), and using \eqref{b=0}, the integration of system \eqref{eq_scale} yields
\begin{equation}\label{sol_invariance_scale}
\begin{aligned}
u_0(x,y)&=x U_0(\omega),\\
v_0(x,y)&=x V_0(\omega),\\
p_0(x,y)&=\kappa_1 \log(x)+P_0(\omega),\\
u_1(x,y)&=x\left((\kappa_2-\kappa_3)U_0(\omega)\log(x)+U_1(\omega)\right),\\
v_1(x,y)&=x\left((\kappa_2-\kappa_3)V_0(\omega)\log(x)+V_1(\omega)\right)\\
&-a_1\left(3 a_3 x^2+2 a_4 x \log(x)-a_5\right),\\
p_1(x,y)&=P_1(\omega)-6 \frac{a_1 a_3}{\Rey}y+\frac{\kappa_1}{2}(\kappa_2-\kappa_3)\log^2(x)\\
&+\left((\kappa_2-\kappa_3)P_0(\omega)-(\kappa_1\kappa_3+6 \frac{a_2 a_3}{\Rey}+a_7)\right)\log(x),
\end{aligned}
\end{equation}
where $\omega=y/x$, and $U_0(\omega)$, $V_0(\omega)$, $P_0(\omega)$, $U_1(\omega)$, $V_1(\omega)$, 
$P_1(\omega)$ are functions to be determined.

Substituting relations \eqref{sol_invariance_scale} into system \eqref{creep_eqs}, and separating at the 
various orders of $\varepsilon$, the following reduced system of ordinary differential equations is 
provided:
\begin{equation}
\label{reduced_scale}
\left\{
\begin{aligned}
&V_0^{\prime}-\omega U_0^{\prime}+U_0 = 0,\\
&(\omega^2+1)U_0^{\prime\prime}+\Rey(\omega P_0^{\prime}-\kappa_1)=0,\\
&\omega(U_0^{\prime\prime}+\omega V_0^{\prime\prime})-\Rey P_0^{\prime}=0,\\
&V_1^{\prime}-\omega U_1^{\prime}+U_1+(\kappa_2-\kappa_3) U_0=0,\\
&(\omega^2+1)\left((\omega U_0-V_0) U_0^{\prime\prime\prime}-\frac{U_1^{\prime\prime}}{\Rey}\right)+
\omega^2 \left(2(\omega V_0^{\prime}-V_0)-U_0^{\prime}\right)V_0^{\prime\prime}\\
&-\omega \left(2(\omega U_0+V_0)-(5\omega^2+2)U_0^{\prime}\right)U_0^{\prime\prime}\\
&+(\kappa_2-\kappa_3)\left(P_0-\frac{1}{\Rey}(U_0-2\omega U_0^{\prime})\right)-\kappa_1\kappa_3\\
&-6 \frac{a_2 a_3}{\Rey}-a_7-\omega P_1^{\prime}=0,\\
&\omega\left(\omega U_0-(\omega^2+1) V_0\right)U_0^{\prime\prime\prime}+\omega^3 U_0 V_0^{\prime\prime
\prime}-\frac{\omega}{\Rey}(\omega V_1^{\prime\prime}+U_1^{\prime\prime})\\
&-\left((5\omega^2+2)U_0^{\prime}+\omega((\omega^2-1)V_0^{\prime}-7 
U_0)+2(\omega^2+1)V_0\right)U_0^{\prime\prime}\\
&-\omega^2(\omega U_0^{\prime}-4 U_0)V_0^{\prime\prime}+\frac{\kappa_2-\kappa_3}{\Rey}(U_0^{\prime}
+2\omega V_0^{\prime}-V_0)\\
&+2\frac{a_1a_4}{\Rey}+P_1^{\prime}=0,
\end{aligned}
\right.
\end{equation}
the prime $^\prime$ denoting the differentiation with respect to $\omega$.

By inserting the solution of system \eqref{reduced_scale} into equations \eqref{sol_invariance_scale}, and 
then into the perturbation expansions \eqref{expand_variables}, 
we get the following approximately invariant solution to system \eqref{creep_eqs}:
\begin{align}\label{solution_scale_1}
u(x,y)&=\frac{\kappa_1 \Rey }{2}y \arctan(y/x) + c_2 x+c_1 y\\
&+\varepsilon\left(\left(\frac{\kappa_2-
\kappa_3}{2}\left(\left(\kappa_1\left(\frac{\log((y/x)^2+1)}{2}+\log(x)\right)+c_4\right)\Rey\, y\right.
\right.\right.\nonumber
\displaybreak[0]\\
&+\left.(c_1+c_3)x-2 c_2 y\phantom{\frac{}{}}\right)-a_1a_4 x\nonumber\displaybreak[0]\\
&-\left.\left(3 a_2 a_3+\frac{\Rey}{2}(\kappa_1\kappa_3+a_7)\right)y\right)\arctan(y/x)\nonumber
\displaybreak[0]\\
&+(\kappa_2-\kappa_3)
(c_2 x+c_1 y)\left(\frac{\log((y/x)^2+1)}{2}+\log(x)\right)\nonumber\displaybreak[0]\\
&+c_6 x+\left.\left(c_5-c_1(\kappa_2-\kappa_3)\right)y\phantom{\frac{}{}}\right),\nonumber\displaybreak[0]
\\
\label{solution_scale_2}
v(x,y)&=-\frac{\kappa_1 \Rey}{2}x \arctan(y/x)+\left(\frac{\kappa_1 \Rey}{2}-
c_2\right) y + c_3 x\displaybreak[0]\\
&+\varepsilon\left(\left(-\frac{\kappa_2-\kappa_3}{2}
\left(\left(\kappa_1\left(\frac{\log((y/x)^2+1)}{2}
+\log(x)\right)+c_4\right)\Rey\, x\right.\right.\right.\displaybreak[0]\nonumber\\
&+2 c_2 x+\left.\left.(c_1+c_3)y\phantom{\frac{}{}}\right)+a_1a_4 y\right.\displaybreak[0]\nonumber\\
&+\left.\left(3 a_2 a_3+\frac{\Rey}{2}(\kappa_1\kappa_2+a_7)\right)x\right)
\arctan(y/x)\displaybreak[0]\nonumber\displaybreak[0]\\
&+\left((\kappa_2-\kappa_3)
\left(c_3 x+\left(\frac{\kappa_1 \Rey}{2}-c_2\right)y\right)\right.\displaybreak[0]\nonumber\\
&-\left.\phantom{\frac{}{}}2 a_1 a_4 x\right)\left(\frac{\log((y/x)^2+1)}{2}+\log(x)\right)-3 a_1 a_3 
x^2+c_7 x
\displaybreak[0]\nonumber\\
&-\left.\left(c_6-\frac{\Rey}{2}(c_4(\kappa_2-\kappa_3)-\kappa_1\kappa_2-a_7)+3a_2 a_3\right) y+a_1 a_5 
\right),\nonumber\displaybreak[0]\\
\label{solution_scale_3}
p(x,y)&=\kappa_1\left(\frac{\log((y/x)^2+1)}{2}+\log(x)\right)+c_4\displaybreak[0]\\
&+\varepsilon\left(-\frac{\kappa_1}{2}(\kappa_2-\kappa_3)
\arctan^2(y/x)+\frac{\kappa_1}{8}(\kappa_2-\kappa_3)\log^2((y/x)^2+1)\right.\displaybreak[0]\nonumber\\
&+\frac{(\kappa_2-\kappa_3)(c_3-c_1)-2 a_1 a_4}
{\Rey}\arctan(y/x)\nonumber\displaybreak[0]\\
&+\left(\left(\frac{\kappa_2-\kappa_3}{2}(\kappa_1 \log(x)+c_4)-\frac{\kappa_1\kappa_3}{2}\right)\right.
\nonumber\displaybreak[0]\\
&-\left.3\frac{a_2 a_3}{\Rey}-\frac{a_7}{2}\right)\log((y/x)^2+1)+\frac{\kappa_1}{2}(\kappa_2-\kappa_3)
\log^2(x)\nonumber\displaybreak[0]\\
&+\left(c_4(\kappa_2-\kappa_3)-\kappa_1\kappa_3-6\frac{a_2 a_3}{\Rey}-a_7\right)\log(x)\nonumber
\displaybreak[0]\\
&+\left.\kappa_1 \Rey\frac{x}{x^2+y^2}\left((4 c_2-\kappa_1 \Rey)x+2(c_1+c_3)y\right)-6\frac{a_1 a_3}{\Rey}
y+c_8\right),\nonumber
\displaybreak[0]
\end{align}
where $c_i$ $(i=1,\ldots,8)$ are arbitrary constants.

\subsection{Approximately invariant solutions with respect to $\Xi_B$}
Let us now consider the approximate generator 
\begin{equation}\label{trasl_oper}
\begin{aligned}
\Xi_B&=(1+\varepsilon \kappa_3)\frac{\partial}{\partial x}+(\kappa_1+\varepsilon \kappa_4)\frac{\partial}
{\partial y}+\varepsilon\frac{\partial^2 f_1}{\partial y^2}\frac{\partial}{\partial u}-\varepsilon
\frac{\partial^2 f_1}{\partial 
x \partial y}\frac{\partial}{\partial v}\\
&+\left(\kappa_2+\varepsilon\left(f_2-\frac{1}{\Rey}
\left(\frac{\partial^3 f_1}{\partial x \partial y^2}+\frac{\partial^3 f_1}{\partial x^3}\right)\right)
\right)\frac{\partial}
{\partial p}.
\end{aligned}
\end{equation}

The corresponding approximately invariant solutions  are such that
\begin{equation}
\begin{aligned}
&(1+\varepsilon \kappa_3)\frac{\partial u}{\partial x}+(\kappa_1+\varepsilon \kappa_4)\frac{\partial u}
{\partial y}=\varepsilon\frac{\partial^2 f_1}{\partial y^2},\\
&(1+\varepsilon \kappa_3)\frac{\partial v}{\partial x}+(\kappa_1+\varepsilon \kappa_4)\frac{\partial v}
{\partial y}=-\varepsilon\frac{\partial^2 f_1}{\partial x \partial y},\\
&(1+\varepsilon \kappa_3)\frac{\partial p}{\partial x}+(\kappa_1+\varepsilon \kappa_4)\frac{\partial p}
{\partial y}=\kappa_2+\varepsilon\left(f_2-\frac{1}{\Rey}
\left(\frac{\partial^3 f_1}{\partial x \partial y^2}+\frac{\partial^3 f_1}{\partial x^3}\right)\right),\\
\end{aligned}
\end{equation}
whereupon, insertion of \eqref{expand_variables}, and separation of the coefficients of different powers 
of $\varepsilon$, provide the system
\begin{equation}\label{eq_trasl}
\left\{\begin{aligned}
&\frac{\partial u_0}{\partial x}+\kappa_1\frac{\partial u_0}{\partial y}=0,\\
&\frac{\partial u_1}{\partial x}+\kappa_1\frac{\partial u_1}{\partial y}+\kappa_3\frac{\partial u_0}
{\partial x}+\kappa_4\frac{\partial u_0}{\partial y}=\frac{\partial^2 f_1}{\partial y^2},\\
&\frac{\partial v_0}{\partial x}+\kappa_1\frac{\partial v_0}{\partial y}=0,\\
&\frac{\partial v_1}{\partial x}+\kappa_1\frac{\partial v_1}{\partial y}+\kappa_3\frac{\partial v_0}
{\partial x}+\kappa_4\frac{\partial v_0}{\partial y}=-\frac{\partial^2 f_1}{\partial x \partial y},\\
&\frac{\partial p_0}{\partial x}+\kappa_1\frac{\partial p_0}{\partial y}=\kappa_2,\\
&\frac{\partial p_1}{\partial x}+\kappa_1\frac{\partial p_1}{\partial y}+\kappa_3\frac{\partial p_0}
{\partial x}+\kappa_4\frac{\partial p_0}{\partial y}=f_2-\frac{1}{\Rey}
\left(\frac{\partial^3 f_1}{\partial x \partial y^2}+\frac{\partial^3 f_1}{\partial x^3}\right).
\end{aligned}
\right.
\end{equation}
We are able to explicitly determine approximately invariant solutions in all the cases (i), (ii) and (iii).

\subsubsection{Case (i)}
In this case, the solution to system \eqref{eq_trasl} is
\begin{equation}\label{sol_trasl_b=0}
\begin{aligned}
u_0(x,y)&=U_0(\omega),\qquad v_0(x,y)=V_0(\omega),\qquad p_0(x,y)=\kappa_2 x +P_0(\omega),\\
u_1(x,y)&=(\kappa_1\kappa_3-\kappa_4)x U_0^{\prime}(\omega)+U_1(\omega),\\
v_1(x,y)&=\left((\kappa_1\kappa_3-\kappa_4)V_0^{\prime}(\omega)-a_1(a_3x^2+a_4x+a_5)\right)x+V_1(\omega),\\
p_1(x,y)&=\left((\kappa_1\kappa_3-\kappa_4)P_0^{\prime}(\omega)-3\frac{a_3}{\Rey}(2(a_1 \omega+a_2)+
\kappa_1 a_1 x)\right.\\
&-\left.\kappa_2\kappa_3 -a_7\phantom{\frac{}{}}\right)x+P_1(\omega),\\
\end{aligned}
\end{equation}
the prime $^\prime$ denoting the differentiation with respect to $\omega$, where $\omega=y-\kappa_1 x$, and 
$U_0(\omega)$, $V_0(\omega)$, $P_0(\omega)$, $U_1(\omega)$, $V_1(\omega)$, $P_1(\omega)$ satisfy the 
following reduced system of ordinary differential equations:
\begin{equation}
\label{reduced_trasl}
\left\{
\begin{aligned}
&V_0^{\prime}-\kappa_1 U_0^{\prime}=0,\\
&(\kappa_1^2+1)U_0^{\prime\prime}+\Rey(\kappa_1 P_0^{\prime}-\kappa_2)=0,\\
&\kappa_1(U_0^{\prime\prime}+\kappa_1 V_0^{\prime\prime})-\Rey P_0^{\prime}=0,\\
&V_1^{\prime}-\kappa_1 U_1^{\prime}+(\kappa_1\kappa_3-\kappa_4)U_0^{\prime}=0,\\
&(\kappa_1^2+1)\left((\kappa_1 U_0-V_0) U_0^{\prime\prime\prime}-\frac{U_1^{\prime\prime}}{\Rey}\right)
+\kappa_1^2 \left(2\kappa_1 V_0^{\prime}-U_0^{\prime}\right)V_0^{\prime\prime}\\
&+\kappa_1\left((5 \kappa_1^2+2)U_0^{\prime}+\frac{2(\kappa_1\kappa_3-\kappa_4)}{\Rey}\right)U_0^{\prime
\prime}-\kappa_1 
P_1^{\prime}\\
&+(\kappa_1\kappa_3-\kappa_4)P_0^{\prime}-\kappa_2\kappa_3-a_7+\alpha=0,\\
&\kappa_1\left(\kappa_1 U_0-(\kappa_1^2+1) V_0\right)U_0^{\prime\prime\prime}+\kappa_1^3 U_0 V_0^{\prime
\prime\prime}-
\frac{\kappa_1}{\Rey}(U_1^{\prime\prime}+\kappa_1 V_1^{\prime\prime})\\
&-\left(\kappa_1(\kappa_1^2-1)V_0^{\prime}+(5 \kappa_1^2+2)U_0^{\prime}-\frac{\kappa_1 \kappa_3-\kappa_4}
{\Rey}\right)U_0^{\prime
\prime}\\
&-\kappa_1\left(\kappa_1^2 U_0^{\prime}-\frac{2}{\Rey}(\kappa_1\kappa_3-\kappa_4)\right)V_0^{\prime\prime}
+P_1^{\prime}+\beta=0,
\end{aligned}
\right.
\end{equation}
where $\alpha=-6 \dfrac{a_3}{\Rey}(a_1 \omega+a_2)$ and $\beta=2\dfrac {a_1a_4}{\Rey}$. 

By inserting the solution of system  \eqref{reduced_trasl} into equations \eqref{sol_trasl_b=0} and then 
into the pertubation expansions \eqref{expand_variables}, the approximately invariant solution of system 
\eqref{creep_eqs} is given:
\begin{align}
u(x,y)&=\frac{\kappa_2 \Rey}{2(\kappa_1^2+1)^{2}}(y-\kappa_1 x)^2+c_1(y-\kappa_1 x)+c_2\\
&+\varepsilon\left(-\frac{a_1 a_3}{(\kappa_1^2+1)^{2}}(y-\kappa_1 x)^{3}+
\left(\frac{\kappa_2\Rey(\kappa_1(3\kappa_1\kappa_3-4\kappa_4)-\kappa_3)}{2(\kappa_1^2+1)^{3}}\right.
\right.\displaybreak[0]\nonumber\\
&-\left.\frac{a_7 \Rey +2(3 a_2 a_3-a_1 a_4\kappa_1)}{2(\kappa_1^2+1)^{2}}\right)(y-\kappa_1 
x)^2\displaybreak[0]\nonumber\\
&+\left.\left(\frac{\kappa_2 \Rey(\kappa_1 \kappa_3-\kappa_4)}{(\kappa_1^2+1)^2}x+c_5\right)(y-\kappa_1 x)
+c_1(\kappa_1\kappa_3-\kappa_4)x+c_6\right),\displaybreak[0]\nonumber\\
v(x,y)&=\frac{\kappa_1\kappa_2 \Rey}{2(\kappa_1^2+1)^{2}}(y-\kappa_1 x)^2+\kappa_1 c_1(y-\kappa_1 x)+c_3\\
&+\varepsilon\left(-\frac{a_1 a_3\kappa_1}{(\kappa_1^2+1)^{2}}(y-\kappa_1 x)^{3}+\left(\frac{\kappa_2 
\Rey(\kappa_4+\kappa_1(2(\kappa_1^2-1)\kappa_3-3\kappa_1\kappa_4))}{2(\kappa_1^2+1)^{3}}\right.\right.
\displaybreak[0]\nonumber\\
&-\left.\frac{\kappa_1(a_7\Rey+2(3a_2a_3-a_1a_4\kappa_1))}{2(\kappa_1^2+1)^{2}}\right)(y-
\kappa_1 x)^2\displaybreak[0]\nonumber\\
&+\left((\kappa_1\kappa_3-\kappa_4)\left(\frac{\kappa_1\kappa_2 \Rey}{(\kappa_1^2+1)^2}x-c_1\right)+
\kappa_1 c_5\right)(y-\kappa_1 x)\displaybreak[0]\nonumber\\
&-\left.\phantom{\frac{}{}}a_1(a_3x^3+a_4x^2)+\left(\kappa_1 c_1(\kappa_1\kappa_3-\kappa_4)-a_1a_5\right) x
+c_7\right),\displaybreak[0]\nonumber\\
p(x,y)&=\frac{\kappa_1\kappa_2}{\kappa_1^2+1}(y-\kappa_1 x)+\kappa_2 x+c_4\\
&+\varepsilon\left(\frac{\kappa_2^2 \Rey^2}
{(\kappa_1^2+1)^{2}}(y-\kappa_1 x)^2+\left(\kappa_2\left(2 c_1 \Rey+
\frac{\kappa_4\kappa_1(\kappa_1\kappa_4+2\kappa_3)}
{(\kappa_1^2+1)^{2}}\right)\right.\right.\displaybreak[0]\nonumber\\
&-\left.\frac{a_7\kappa_1\Rey +2(3a_2a_3\kappa_1+a_1a_4)}{\Rey(\kappa_1^2+1)}\right)(y-\kappa_1 x)
\displaybreak[0]\nonumber\\
&+\frac{3a_1 a_3}{\Rey(\kappa_1^2+1)}(\kappa_1(x^2-y^2)-2xy)\displaybreak[0]\nonumber\\
&-\left.\left(a_7+\frac{6a_2 a_3}{\Rey}+\frac{\kappa_2(\kappa_1\kappa_4+\kappa_3)}{\kappa_1^2+1}\right)x
+c_8\right),\displaybreak[0]\nonumber
\end{align}
where $c_i$ $(i=1,\ldots,8)$ are arbitrary constants.

\subsubsection{Case (ii)}
In this case, the representation of the approximately invariant solution is

\begin{align}\label{sol_trasl_b>0}
u_0(x,y)&=U_0(\omega),\qquad v_0(x,y)=V_0(\omega),\qquad p_0(x,y)=\kappa_2 x +P_0(\omega),\displaybreak[0]\nonumber\\
u_1(x,y)&=\left(\frac{a_2\exp{(-by)}(a_5-a_6 \kappa_1)+a_1\exp{(by)}(a_5+a_6 \kappa_1)}{\kappa_1^2+1}b x
\right.\displaybreak[0]\nonumber\\
&+a_2\exp{(-by)}\frac{b(a_3-a_4\kappa_1)(\kappa_1^2+1)-a_6(\kappa_1^2-1)+2a_5\kappa_1}{(\kappa_1^2+1)^2}\displaybreak[0]\nonumber\\
&+a_1\exp{(by)}\left.\frac{b(a_3+a_4\kappa_1)(\kappa_1^2+1)-a_6(\kappa_1^2-1)-2a_5\kappa_1}
{(\kappa_1^2+1)^2}\right)\sin(b x)\displaybreak[0]\nonumber\\
&-\left(\frac{a_2\exp{(-by)}(a_6+a_5 \kappa_1)+a_1\exp{(by)}(a_6-a_5 \kappa_1)}{\kappa_1^2+1}b x\right.\displaybreak[0]\nonumber\\
&+a_2\exp{(-by)}\frac{b(a_4+a_3\kappa_1)(\kappa_1^2+1)+a_5(\kappa_1^2-1)+2a_6\kappa_1}{(\kappa_1^2+1)^2}\displaybreak[0]\nonumber\\
&+a_1\exp{(by)}\left.\frac{b(a_4-a_3\kappa_1)(\kappa_1^2+1)+a_5(\kappa_1^2-1)-2a_6\kappa_1}
{(\kappa_1^2+1)^2}\right)\cos(b x)\displaybreak[0]\nonumber\\
&+(\kappa_1\kappa_3-\kappa_4) x U_0^{\prime}(\omega)+U_1(\omega),\displaybreak[0]\nonumber\\
v_1(x,y)&=\left(\frac{a_2\exp{(-by)}(a_6+a_5 \kappa_1)-a_1\exp{(by)}(a_6-a_5 \kappa_1)}{\kappa_1^2+1}b x
\right.\displaybreak[0]\nonumber\\
&+a_2\exp{(-by)}\frac{b(a_4+a_3\kappa_1)(\kappa_1^2+1)-a_6\kappa_1(\kappa_1^2-1)+2a_5\kappa_1^2}
{(\kappa_1^2+1)^2}\displaybreak[0]\nonumber\\
&-a_1\exp{(by)}\left.\frac{b(a_4-a_3\kappa_1)(\kappa_1^2+1)+a_6\kappa_1(\kappa_1^2-1)+2a_5\kappa_1^2}
{(\kappa_1^2+1)^2}\right)
\sin(b x)\displaybreak[0]\nonumber\\
&+\left(\frac{a_2\exp{(-by)}(a_5-a_6\kappa_1)-a_1\exp{(by)}(a_5+a_6 \kappa_1)}{\kappa_1^2+1}b x\right.\displaybreak[0]\nonumber\\
&+a_2\exp{(-by)}\frac{b(a_3-a_4\kappa_1)(\kappa_1^2+1)-a_5\kappa_1(\kappa_1^2-1)-2a_6\kappa_1^2}
{(\kappa_1^2+1)^2}\displaybreak[0]\nonumber\\
&-a_1\exp{(by)}\left.\frac{b(a_3+a_4\kappa_1)(\kappa_1^2+1)+a_5\kappa_1(\kappa_1^2-1)-2a_6\kappa_1^2}
{(\kappa_1^2+1)^2}\right)
\cos(b x)\displaybreak[0]\nonumber\\
&+(\kappa_1\kappa_3-\kappa_4) x V_0^{\prime}(\omega)+V_1(\omega),\displaybreak[0]\nonumber\\
p_1(x,y)&=\frac{2 b\left(a_2\exp{(-by)}(a_5-a_6 \kappa_1)+a_1\exp{(by)}(a_5+a_6 \kappa_1)\right)}
{\Rey(\kappa_1^2+1)}\sin(b x)\displaybreak[0]\nonumber\\
&-\frac{2 b\left(a_2\exp{(-by)}(a_6+a_5 \kappa_1)+a_1\exp{(by)}(a_6-a_5 \kappa_1)\right)}
{\Rey(\kappa_1^2+1)}\cos(b x)\displaybreak[0]\nonumber\\
&+\left((\kappa_1\kappa_3-\kappa_4) P_0^{\prime}(\omega)-\kappa_2\kappa_3-a_7\right)x+P_1(\omega),
\end{align}
the prime $^\prime$ denoting the differentiation with respect to $\omega$, where $\omega=y-\kappa_1 x$, and 
$U_0(\omega)$, $V_0(\omega)$, $P_0(\omega)$, $U_1(\omega)$, $V_1(\omega)$, $P_1(\omega)$ satisfy the  
reduced system \eqref{reduced_trasl} with $\alpha=\beta=0$.

By solving the reduced system, we finally obtain the following approximately invariant solution:
\begin{align}
u(x,y)&=\frac{\kappa_2 \Rey}{2(\kappa_1^2+1)^{2}}(y-\kappa_1 x)^2+c_1(y-\kappa_1 x)+c_2\\
&+\varepsilon\left(\Rey\frac{\kappa_2(\kappa_1(3\kappa_1\kappa_3-4\kappa_4)-\kappa_3)-a_7(\kappa_1^2+1)}
{2(\kappa_1^2+1)^{3}}(y-\kappa_1 x)^2\right.\displaybreak[0]\nonumber\\
&+\left.\left(\frac{\kappa_2 \Rey(\kappa_1 \kappa_3-\kappa_4)}{(\kappa_1^2+1)^2}x+c_5\right)(y-\kappa_1 x)
\right.\displaybreak[0]\nonumber\\
&+\left(\frac{a_2\exp{(-by)}(a_5-a_6 \kappa_1)+a_1\exp{(by)}(a_5+a_6 \kappa_1)}{\kappa_1^2+1}b x\right.
\displaybreak[0]\nonumber\\
&+a_2\exp{(-by)}\frac{b(a_3-a_4\kappa_1)(\kappa_1^2+1)-a_6(\kappa_1^2-1)+2a_5\kappa_1}{(\kappa_1^2+1)^2}
\displaybreak[0]\nonumber\\
&+a_1\exp{(by)}\left.\frac{b(a_3+a_4\kappa_1)(\kappa_1^2+1)-a_6(\kappa_1^2-1)-2a_5\kappa_1}
{(\kappa_1^2+1)^2}\right)\sin(b x)
\displaybreak[0]\nonumber\\
&-\left(\frac{a_2\exp{(-by)}(a_6+a_5 \kappa_1)+a_1\exp{(by)}(a_6-a_5 \kappa_1)}{\kappa_1^2+1}b x\right.
\displaybreak[0]\nonumber\\
&+a_2\exp{(-by)}\frac{b(a_4+a_3\kappa_1)(\kappa_1^2+1)+a_5(\kappa_1^2-1)+2a_6\kappa_1}{(\kappa_1^2+1)^2}
\displaybreak[0]\nonumber\\
&+a_1\exp{(by)}\left.\frac{b(a_4-a_3\kappa_1)(\kappa_1^2+1)+a_5(\kappa_1^2-1)-2a_6\kappa_1}
{(\kappa_1^2+1)^2}\right)\cos(b x)
\displaybreak[0]\nonumber\\
&+\left.\phantom{\frac{}{}}c_1(\kappa_1\kappa_3-
\kappa_4)x+c_6\right),\displaybreak[0]\nonumber\\
v(x,y)&=\frac{\kappa_1\kappa_2 \Rey}{2(\kappa_1^2+1)^{2}}(y-\kappa_1 x)^2+\kappa_1 c_1(y-\kappa_1 x)+c_3\\
&+\varepsilon\left(\Rey\frac{\kappa_2 (\kappa_4+\kappa_1(2(\kappa_1^2-1)\kappa_3-3\kappa_1\kappa_4))-a_7 
\kappa_1(\kappa_1^2+1)}{2(\kappa_1^2+1)^{3}}(y-\kappa_1 x)^2\right.\displaybreak[0]\nonumber\\
&+\left((\kappa_1\kappa_3-\kappa_4)\left(\frac{\kappa_1\kappa_2 \Rey}{(\kappa_1^2+1)^2}x-c_1\right)+
\kappa_1 c_5\right)(y-\kappa_1 x)\displaybreak[0]\nonumber\\
&+\left(\frac{a_2\exp{(-by)}(a_6+a_5 \kappa_1)-a_1\exp{(by)}(a_6-a_5 \kappa_1)}{\kappa_1^2+1}b x\right.
\displaybreak[0]\nonumber\\
&+a_2\exp{(-by)}\frac{b(a_4+a_3\kappa_1)(\kappa_1^2+1)-a_6\kappa_1(\kappa_1^2-1)+2a_5\kappa_1^2}
{(\kappa_1^2+1)^2}\displaybreak[0]\nonumber\\
&-a_1\exp{(by)}\left.\frac{b(a_4-a_3\kappa_1)(\kappa_1^2+1)+a_6\kappa_1(\kappa_1^2-1)+2a_5\kappa_1^2}
{(\kappa_1^2+1)^2}\right)\sin(b x)\displaybreak[0]\nonumber\\
&+\left(\frac{a_2\exp{(-by)}(a_5-a_6\kappa_1)-a_1\exp{(by)}(a_5+a_6 \kappa_1)}{\kappa_1^2+1}b x\right.
\displaybreak[0]\nonumber\\
&+a_2\exp{(-by)}\frac{b(a_3-a_4\kappa_1)(\kappa_1^2+1)-a_5\kappa_1(\kappa_1^2-1)-2a_6\kappa_1^2}
{(\kappa_1^2+1)^2}\displaybreak[0]\nonumber\\
&-a_1\exp{(by)}\left.\frac{b(a_3+a_4\kappa_1)(\kappa_1^2+1)+a_5\kappa_1(\kappa_1^2-1)-2a_6\kappa_1^2}
{(\kappa_1^2+1)^2}\right)\cos(b x)\displaybreak[0]\nonumber\\
&+\left.\phantom{\frac{}{}}\kappa_1c_1(\kappa_1\kappa_3-\kappa_4)x+c_7\right),\displaybreak[0]\nonumber\\
p(x,y)&=\frac{\kappa_1\kappa_2}{\kappa_1^2+1}(y-\kappa_1 x)+\kappa_2 x+c_4\displaybreak[0]\nonumber\\
&+\varepsilon\left(\frac{\kappa_2^2 \Rey^2}{(\kappa_1^2+1)^{2}}(y-\kappa_1 x)^2+2\kappa_2\left(c_1\Rey -
\frac{\kappa_1\kappa_3-\kappa_4}{(\kappa_1^2+1)^2}\right)(y-\kappa_1 x)\right.\displaybreak[0]\nonumber\\
&+\frac{2 b\left(a_2\exp{(-by)}(a_5-a_6 \kappa_1)+a_1\exp{(by)}(a_5+a_6 \kappa_1)\right)}
{\Rey(\kappa_1^2+1)}\sin(b x)\displaybreak[0]\nonumber\\
&-\frac{2 b\left(a_2\exp{(-by)}(a_6+a_5 \kappa_1)+a_1\exp{(by)}(a_6-a_5 \kappa_1)\right)}
{\Rey(\kappa_1^2+1)}\cos(b x)\displaybreak[0]\nonumber\\
&-\left.\frac{\kappa_2\kappa_3+a_7}{\kappa_1^2+1}x-\frac{\kappa_2\kappa_4+a_7\kappa_1}{\kappa_1^2+1}y
+c_8\right),
\end{align}
where $c_i$ $(i=1,\ldots,8)$ are arbitrary constants.

\subsubsection{Case (iii)}
In this last case, the representation of the approximately invariant solution is
\begin{align}\label{sol_trasl_b<0}
u_0(x,y)&=U_0(\omega),\qquad v_0(x,y)=V_0(\omega),\qquad p_0(x,y)=\kappa_2 x +P_0(\omega),\displaybreak[0]\nonumber\\
u_1(x,y)&=\left(\frac{a_6\exp{(-bx)}(a_2-a_1 \kappa_1)-a_5\exp{(bx)}(a_2+a_1 \kappa_1)}{\kappa_1^2+1}b x
\right.\displaybreak[0]\nonumber\\
&+\exp{(-bx)}\frac{b a_4(a_2-a_1\kappa_1)(\kappa_1^2+1)-a_6(a_2(\kappa_1^2-1)+2a_1\kappa_1)}
{(\kappa_1^2+1)^2}\displaybreak[0]\nonumber\\
&-\left.\exp{(bx)}\frac{b a_3(a_2+a_1\kappa_1)(\kappa_1^2+1)+a_5(a_2(\kappa_1^2-1)-2a_1\kappa_1)}
{(\kappa_1^2+1)^2}\right)\sin(b y)\displaybreak[0]\nonumber\\
&+\left(\frac{a_6\exp{(-bx)}(a_1+a_2 \kappa_1)-a_5\exp{(bx)}(a_1-a_2 \kappa_1)}{\kappa_1^2+1}b x\right.\displaybreak[0]\nonumber\\
&+\exp{(-bx)}\frac{b a_4(a_1+a_2\kappa_1)(\kappa_1^2+1)-a_6(a_1(\kappa_1^2-1)-2a_2\kappa_1)}
{(\kappa_1^2+1)^2}\displaybreak[0]\nonumber\\
&-\left.\exp{(bx)}\frac{b a_3(a_1-a_2\kappa_1)(\kappa_1^2+1)+a_5(a_1(\kappa_1^2-1)+2a_2\kappa_1)}
{(\kappa_1^2+1)^2}\right)\cos(b y)\displaybreak[0]\nonumber\\
&+(\kappa_1\kappa_3-\kappa_4) x U_0^{\prime}(\omega)+U_1(\omega),\displaybreak[0]\nonumber\\
v_1(x,y)&=\left(\frac{a_6\exp{(-bx)}(a_1+a_2 \kappa_1)+a_5\exp{(bx)}(a_1-a_2 \kappa_1)}{\kappa_1^2+1}b x
\right.\displaybreak[0]\nonumber\\
&+\exp{(-bx)}\frac{b a_4(a_1+a_2\kappa_1)(\kappa_1^2+1)-a_6\kappa_1(a_2(\kappa_1^2-1)+2a_1\kappa_1)}
{(\kappa_1^2+1)^2}\displaybreak[0]\nonumber\\
&+\left.\exp{(bx)}\frac{b a_3(a_1-a_2\kappa_1)(\kappa_1^2+1)-a_5\kappa_1(a_2(\kappa_1^2-1)-2a_1\kappa_1)}
{(\kappa_1^2+1)^2}\right)\sin(b y)\displaybreak[0]\nonumber\\
&-\left(\frac{a_6\exp{(-bx)}(a_2-a_1 \kappa_1)+a_5\exp{(bx)}(a_2+a_1 \kappa_1)}{\kappa_1^2+1}b x\right.\displaybreak[0]\nonumber\\
&+\exp{(-bx)}\frac{b a_4(a_2-a_1\kappa_1)(\kappa_1^2+1)+a_6\kappa_1(a_1(\kappa_1^2-1)-2a_2\kappa_1)}
{(\kappa_1^2+1)^2}\displaybreak[0]\nonumber\\
&+\left.\exp{(bx)}\frac{b a_3(a_2+a_1\kappa_1)(\kappa_1^2+1)+a_5\kappa_1(a_1(\kappa_1^2-1)+2a_2\kappa_1)}
{(\kappa_1^2+1)^2}\right)\cos(b y)\displaybreak[0]\nonumber\\
&+(\kappa_1\kappa_3-\kappa_4) x V_0^{\prime}(\omega)+V_1(\omega),\displaybreak[0]\nonumber\\
p_1(x,y)&=\frac{2 b\left(a_6\exp{(-bx)}(a_2-a_1 \kappa_1)-a_5\exp{(bx)}(a_2+a_1 \kappa_1)\right)}
{\Rey(\kappa_1^2+1)}\sin(b y)\displaybreak[0]\nonumber\\
&+\frac{2 b\left(a_6\exp{(-bx)}(a_1+a_2 \kappa_1)-a_5\exp{(bx)}(a_1-a_2 \kappa_1)\right)}
{\Rey(\kappa_1^2+1)}\cos(b y)\displaybreak[0]\nonumber\\
&+\left((\kappa_1\kappa_3-\kappa_4) P_0^{\prime}(\omega)-\kappa_2\kappa_3-a_7\right)x+P_1(\omega),
\end{align}
the prime $^\prime$ denoting the differentiation with respect to $\omega$, where $\omega=y-\kappa_1 x$, and 
$U_0(\omega)$, $V_0(\omega)$, $P_0(\omega)$, $U_1(\omega)$, $V_1(\omega)$, $P_1(\omega)$ satisfy, once 
again, the  reduced system \eqref{reduced_trasl} with $\alpha=\beta=0$.

Finally, we are able to recover the following approximately invariant solution:
\begin{align}
u(x,y)&=\frac{\kappa_2 \Rey}{2(\kappa_1^2+1)^{2}}(y-\kappa_1 x)^2+c_1(y-\kappa_1 x)+c_2\\
&+\varepsilon\left(\Rey\frac{\kappa_2(\kappa_1(3\kappa_1\kappa_3-4\kappa_4)-\kappa_3)-a_7(\kappa_1^2+1)}
{2(\kappa_1^2+1)^{3}}(y-\kappa_1 x)^2\right.\displaybreak[0]\nonumber\\
&+\left.\left(\frac{\kappa_2 \Rey(\kappa_1 \kappa_3-\kappa_4)}{(\kappa_1^2+1)^2}x+c_5\right)(y-\kappa_1 x)
\right.\displaybreak[0]\nonumber\\
&+\left(\frac{a_6\exp{(-bx)}(a_2-a_1 \kappa_1)-a_5\exp{(bx)}(a_2+a_1 \kappa_1)}{\kappa_1^2+1}b x\right.
\displaybreak[0]\nonumber\\
&+\exp{(-bx)}\frac{b a_4(a_2-a_1\kappa_1)(\kappa_1^2+1)-a_6(a_2(\kappa_1^2-1)+2a_1\kappa_1)}
{(\kappa_1^2+1)^2}\displaybreak[0]\nonumber\\
&-\left.\exp{(bx)}\frac{b a_3(a_2+a_1\kappa_1)(\kappa_1^2+1)+a_5(a_2(\kappa_1^2-1)-2a_1\kappa_1)}
{(\kappa_1^2+1)^2}\right)\sin(b y)\displaybreak[0]\nonumber\\
&+\left(\frac{a_6\exp{(-bx)}(a_1+a_2 \kappa_1)-a_5\exp{(bx)}(a_1-a_2 \kappa_1)}{\kappa_1^2+1}b x\right.
\displaybreak[0]\nonumber\\
&+\exp{(-bx)}\frac{b a_4(a_1+a_2\kappa_1)(\kappa_1^2+1)-a_6(a_1(\kappa_1^2-1)-2a_2\kappa_1)}
{(\kappa_1^2+1)^2}\displaybreak[0]\nonumber\\
&-\left.\exp{(bx)}\frac{b a_3(a_1-a_2\kappa_1)(\kappa_1^2+1)+a_5(a_1(\kappa_1^2-1)+2a_2\kappa_1)}
{(\kappa_1^2+1)^2}\right)\cos(b y)\displaybreak[0]\nonumber\\
&+\left.\phantom{\frac{}{}}c_1(\kappa_1\kappa_3-\kappa_4)x+c_6\right),\displaybreak[0]\nonumber\\
v(x,y)&=\frac{\kappa_1\kappa_2 \Rey}{2(\kappa_1^2+1)^{2}}(y-\kappa_1 x)^2+\kappa_1 c_1(y-\kappa_1 x)+c_3\\
&+\varepsilon\left(\Rey\frac{\kappa_2 (\kappa_4+\kappa_1(2(\kappa_1^2-1)\kappa_3-3\kappa_1\kappa_4))-a_7 
\kappa_1(\kappa_1^2+1)}{2(\kappa_1^2+1)^{3}}(y-\kappa_1 x)^2\right.\displaybreak[0]\nonumber\\
&+\left((\kappa_1\kappa_3-\kappa_4)\left(\frac{\kappa_1\kappa_2 \Rey}{(\kappa_1^2+1)^2}x-c_1\right)+
\kappa_1 c_5\right)(y-\kappa_1 x)\displaybreak[0]\nonumber\\
&+\left(\frac{a_6\exp{(-bx)}(a_1+a_2 \kappa_1)+a_5\exp{(bx)}(a_1-a_2 \kappa_1)}{\kappa_1^2+1}b x\right.
\displaybreak[0]\nonumber\\
&+\exp{(-bx)}\frac{b a_4(a_1+a_2\kappa_1)(\kappa_1^2+1)-a_6\kappa_1(a_2(\kappa_1^2-1)+2a_1\kappa_1)}
{(\kappa_1^2+1)^2}\displaybreak[0]\nonumber\\
&+\left.\exp{(bx)}\frac{b a_3(a_1-a_2\kappa_1)(\kappa_1^2+1)-a_5\kappa_1(a_2(\kappa_1^2-1)-2a_1\kappa_1)}
{(\kappa_1^2+1)^2}\right)\sin(b y)\displaybreak[0]\nonumber\\
&-\left(\frac{a_6\exp{(-bx)}(a_2-a_1 \kappa_1)+a_5\exp{(bx)}(a_2+a_1 \kappa_1)}{\kappa_1^2+1}b x\right.
\displaybreak[0]\nonumber\\
&+\exp{(-bx)}\frac{b a_4(a_2-a_1\kappa_1)(\kappa_1^2+1)+a_6\kappa_1(a_1(\kappa_1^2-1)-2a_2\kappa_1)}
{(\kappa_1^2+1)^2}\displaybreak[0]\nonumber\\
&+\left.\exp{(bx)}\frac{b a_3(a_2+a_1\kappa_1)(\kappa_1^2+1)+a_5\kappa_1(a_1(\kappa_1^2-1)+2a_2\kappa_1)}
{(\kappa_1^2+1)^2}\right)\cos(b y)\displaybreak[0]\nonumber\\
&+\left.\phantom{\frac{}{}}\kappa_1c_1(\kappa_1\kappa_3-\kappa_4)x+c_7\right),\displaybreak[0]\nonumber\\
p(x,y)&=\frac{\kappa_1\kappa_2}{\kappa_1^2+1}(y-\kappa_1 x)+\kappa_2 x+c_4\\
&+\varepsilon\left(\frac{\kappa_2^2 \Rey^2}{(\kappa_1^2+1)^{2}}(y-\kappa_1 x)^2+2\kappa_2\left(c_1\Rey -
\frac{\kappa_1\kappa_3-\kappa_4}{(\kappa_1^2+1)^2}\right)(y-\kappa_1 x)\right.\displaybreak[0]\nonumber\\
&+\frac{2 b\left(a_6\exp{(-bx)}(a_2-a_1 \kappa_1)-a_5\exp{(bx)}(a_2+a_1 \kappa_1)\right)}
{\Rey(\kappa_1^2+1)}\sin(b y)\displaybreak[0]\nonumber\\
&+\frac{2 b\left(a_6\exp{(-bx)}(a_1+a_2 \kappa_1)-a_5\exp{(bx)}(a_1-a_2 \kappa_1)\right)}
{\Rey(\kappa_1^2+1)}\cos(b y)\displaybreak[0]\nonumber\\
&-\left.\frac{\kappa_2\kappa_3+a_7}{\kappa_1^2+1}x-\frac{\kappa_2\kappa_4+a_7\kappa_1}{\kappa_1^2+1}y
+c_8\right),\displaybreak[0]\nonumber
\end{align}
where $c_i$ $(i=1,\ldots,8)$ are arbitrary constants.

\subsection{A boundary value problem}
In \cite{Pakdemirli2001}, the physical problem of a mud flow over a porous surface has been considered. 
According to this model, the porosity and the suction velocity increases over the length. In nature,
a mud flow, that occurs especially on the slopes surrounding young, narrow and asymmetric depression 
basins, may spread on detrital porous sediments starting from less porous sandy parts to more porous 
gravelly parts of the plain. In this situation, the boundary conditions read
\begin{equation}
\begin{aligned}\label{boundary_exact}
&u(x,0)=0,\qquad v(x,0)=-v_0 x,\qquad u(-\infty,y)=u_0 y,\\
&\frac{\partial v}{\partial y}(x,\infty)=0,\qquad p(-\infty,y)=p_0,
\end{aligned}
\end{equation}
and the solution has been explicitly given in \cite{Pakdemirli2001}.

Here, we want to analyze the same boundary value problem, but in the approximate sense. 
Let us consider the approximate boundary value problem
\begin{equation}\label{boundary_approx}
\begin{aligned}
&u(x,0)=O(\varepsilon),\quad v(x,0)=-v_0 x+O(\varepsilon),\quad u(-\infty,y)=u_0 y+O(\varepsilon),\\
&\frac{\partial v}{\partial y}(x,\infty)=O(\varepsilon),\qquad p(-\infty,y)=p_0+O(\varepsilon).
\end{aligned}
\end{equation}
By using \eqref{boundary_approx} into the approximately invariant solution given by
\eqref{solution_scale_1}--\eqref{solution_scale_3}, we obtain the 
approximate solution:
\begin{equation}\label{sol_boundary}
\begin{aligned}
u(x,y)&=u_0 y+\varepsilon\left(-a_1 a_4 x \arctan(y/x)+c_5 y\right),\\
v(x,y)&=-v_0 x+\varepsilon\left(-3 a_1 a_3 x^2+c_7 x+a_1 a_5\right.\\
&+a_1 a_4\left.\left(y \arctan(y/x)-x \log(x^2+y^2)\right)\right),\\
p(x,y)&=p_0+\varepsilon\left(-2\frac{a_1 a_4}{\Rey}\arctan(y/x)-6\frac{a_1 a_3}{\Rey}y+c_8\right).
\end{aligned}
\end{equation}
It is immediately to be verified that the exact solution given in \cite{Pakdemirli2001} can be recovered in 
\eqref{sol_boundary} for $\varepsilon=0$.  

\section{Conclusions}
\label{sec:conclusions}
In this paper, we explicitly determine some classes of approximately invariant solutions of the steady creeping  
flow equations of second grade  fluids by using a recently introduced  approach \cite{DSGO-2018} to 
approximate Lie symmetries that is consistent with the principles of perturbative analysis. The same 
equations have been analyzed in \cite{DolapciPakdemirli2004}, where the results of three different 
approximate symmetry methods have been compared. In \cite{DolapciPakdemirli2004}, the authors show that some solutions
can be obtained with one method (essentially, the Fushchich--Shtelen method and, \emph{a fortiori}, with their method, that
simply shorten the length of the needed computations but is not general, since it is assumed that the differential equations are linear at zeroth--order), but not with the Baikov--Gazizov--Ibragimov method. The same problem is not encountered here;  
in all cases approximately invariant solutions can be 
determined and often are more general of the ones characterized in \cite{DolapciPakdemirli2004}.

The method here used, proposed along the lines of the approach by Baikov--Gazizov--Ibragimov 
\cite{BGI-1989}, but taking into account the expansion of the dependent variables,  allows to yield
correct terms for the approximate solutions.
Further applications of the approach proposed in \cite{DSGO-2018} are currently under investigation, and 
aim to show the advantages of the method when analyzing the approximate symmetries of differential 
equations containing small terms.  

\section*{Acknowledgments} 
Work supported by G.N.F.M. of the ``Istituto Nazionale di Alta Matematica F. Severi''. 
The author thanks the unknown Referees for the helpful comments leading to clarify some aspects and improve the
paper.

\end{document}